# The quest for high critical current in applied high-temperature superconductors

Andreas Glatz, Ivan A. Sadovskyy, Ulrich Welp, Wai-Kwong Kwok, and George W. Crabtree

Argonne National Laboratory, Lemont, USA

We present a perspective on a new critical-current-by-design paradigm to tailor and enhance the current-carrying capacity of applied superconductors. Critical current by design is based on large-scale simulations of vortex matter pinning in high-temperature superconductors and has qualitative and quantitative predictive powers to elucidate vortex dynamics under realistic conditions and to propose vortex pinning defects that could enhance the critical current, particularly at high magnetic fields. The simulations are validated with controlled experiments and demonstrate a powerful tool for designing high-performance superconductors for targeted applications.

## Introduction

The widespread application of high-temperature superconducting (HTS) materials is currently at a watershed moment. The desire for higher performing, lower cost superconductors continues to grow with new applications in light-weight motors and generators, the electricity grid of the future, high-energy accelerators for X-ray sources and particle colliders, and potential compact fusion systems. Prior to the discovery of HTS, in the pioneering phase of applied superconductivity, the widespread practical application of type-II superconductors was envisioned by Ted Geballe and John Hulm [1], despite the challenging requirement for helium-cooling of superconducting materials at that time.

The key to enhancing the current carrying capacity of superconductors at high magnetic fields is controlling vortex matter, a collection of single quantum, magnetic flux derived from circulating supercurrents whose behavior determines the entire electromagnetic response of superconducting materials. The driven motion of these vortices due to the Lorentz force in the presence of a magnetic field and applied current induces a finite voltage, thereby negating the desirable zero-resistance feature of a superconductor. Since the core of the vortex is in the normal state, where the superconducting order parameter is zero [see Fig. 1(a)], a solution to mitigate dissipation is to introduce microscopic non-superconducting defects that could 'pin' the vortices, as they would like to sit ontop of such defects to conserve energy, see Figs. 1(b) and 1(c). Efforts to discover the ideal or optimum 'pinning centers' to mitigate vortex motion for targeted field/current applications have dominated the fundamental as well as applied research on enhancing the current carrying capacity of superconductors.

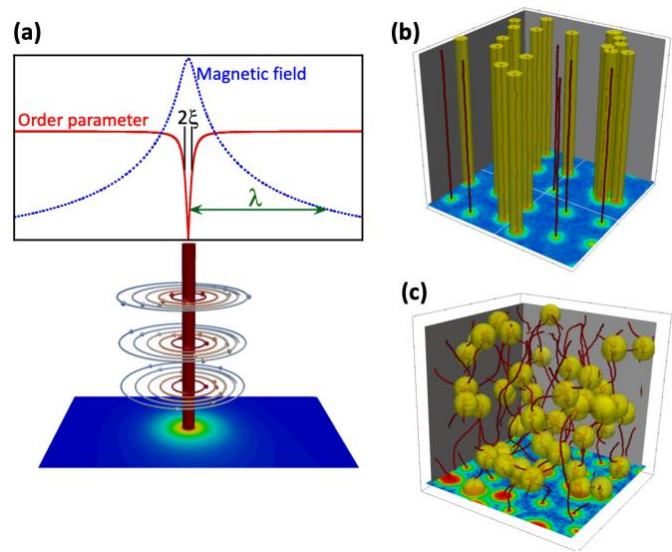

Figure 1. (a) Characteristics of an isolated vortex line: (top) radial distributions of the magnetic field and order parameter amplitude; (bottom) visualized vortex line (red) with circulating currents, screening the magnetic flux. (b) Pinning of vortex lines (red) by columnar defects (yellow). Pinned vortex lines get trapped while free vortices can move due to Lorentz forces (snapshot of a TDGL simulation). (c) As in (b), but with spherical pinning defects (yellow).

In recent years, a new approach to creating high-performance superconducting materials based on self-assembled artificial pinning centers, particle irradiation, and *quantitative predictions* of the critical current, $J_c$, — the largest current the superconductor can carry without losses — via large-scale simulations has gained traction and transformed the way we think about defect synthesis and vortex pinning. This approach, embodied in the new *critical-current-by-design* paradigm [2, 3], heralds a rational and predictive approach to enhancing the current carrying capacity of high-temperature superconductors for targeted applications.

The critical-current-by-design paradigm is based on the combination of three frontier developments in superconductivity: (i) creation of point, line, and cluster pinning sites by post-synthesis ion irradiation, (ii) quantitative prediction of vortex dynamics and critical currents in arbitrary pinning landscapes by large-scale simulations, and (iii) computation-guided chemical synthesis of strong pinning self-assembled nanorods embedded in superconducting materials. In particular, large-scale simulations can reveal in rich detail, the quantitative link between complex nanoscopic pinning landscapes and the resultant macroscopic critical currents. Especially, the route to achieving greater vortex pinning by defects leading to higher critical current densities, $J_c$, is highly nontrivial due to (i) long-range repulsive vortex-vortex interactions, (ii) vortex-defect attractions, and (iii) Lorentz forces on vortices stemming from local currents. The combination of these effects leads to complex collective vortex dynamics, which, in particular, reveals the competing interactions among diverse isotropic and anisotropic defect types [4].

Computer simulations that advance critical-current-by-design should take into account all the above effects and be simple enough so that large systems can be simulated. There are two standard models that satisfy these conditions: Langevin dynamics where each vortex is considered as a point in 2D or an elastic string in 3D [5] and time-dependent Ginzburg-Landau (TDGL) equations for the superconducting order parameter [2]. The first approach is simpler, but misses the intricate collective interaction of vortices, while the second one is more accurate and takes into account the complete vortex matter dynamics. The use of modern computing technologies enables the simulation of large enough systems to capture bulk vortex behavior [6] — e.g., currently sizes up to $(25,000\ \xi)^2$ in two dimensions (2D) or $(800\ \xi)^3$ in three dimensions (3D) within the TDGL framework on a state-of-the-art graphics processing unit. Here $\xi$ is the superconducting coherence length, which defines the typical size of a vortex and at the same time the characteristic length scale of the TDGL equations. In a typical HTS at 77 K, the in-plane coherence length is $\xi \sim 4$ nm, which means that mesocopic sizes of $\sim (100\ \mu m)^2$ [2D] or $\sim (4\ \mu m)^3$ [3D] can be simulated. In a typical simulation, $\xi$ is discretized by two mesh point per dimension, which translates into a computational mesh with up to $10^9$ mesh points for the aforementioned system sizes. This model underlies most of the simulation results discussed in this article. Large-scale simulation of the TDGL equations can elucidate the collective interactions within large arrays of flexible moving vortices and complex pinning sites that produce a single characteristic critical current. Moreover, it can predict the defect structures that can induce extremely high $J_c$, which will lower the cost/performance metric of superconducting wires, and provide a game-changing toolbox that can unlock the potential for all superconductors.

# Vortices and critical currents

The quest for higher critical current in superconductors is often carried out via laborious experimental trial and error. For example, it is known that vortex pinning and hence the critical current in cuprate high-temperature superconductors such as $YBa_2Cu_3O_{7-x}$ (short YBCO) can be enhanced with the addition of rare earth elements (short RE, such as Dy, Ho, etc.) that form insulating nanoparticle precipitates which may self-assemble into either isolated nanoparticles or arrays of columnar rods. However, there is no fundamental understanding of the mechanisms governing the pinning of large vortex arrays by a distribution of such defects, nor is there a universal quantitative model predicting the critical current performance of a given pinning defect array. This conspicuous gap in our fundamental understanding is a major impediment to rapid progress in improving the performance of superconducting wires for applications, see Fig 2.

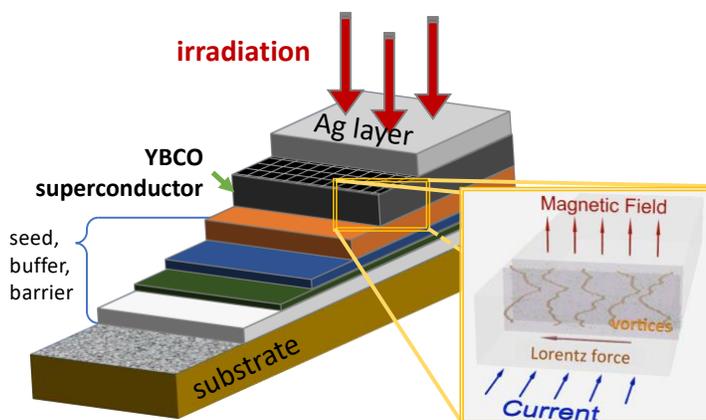

Figure 2. Schematics of a second-generation superconducting wire, an $YBa_2Cu_3O_{7-x}$ coated conductor with complex multilayer architecture. These conductors can be irradiated to improve critical currents. Inset shows vortex dynamics in superconductor with defects due to Lorentz forces.

The dynamics of large-scale arrays of vortices, which determine fundamental superconducting features such as the critical current, $J_c$, remain poorly understood. The reason is the complexity of the vortex system, consisting of flexible vortex lines that interact with each other and with arrays of pinning defects in the composition and structure of the host superconductor. These pinning defects retard or fully halt the motion of vortices, the basic requirement for achieving current flow without energy loss. While the behavior of a single vortex interacting with a single pinning defect is fairly well understood, the transition to large numbers of vortices and pinning defects introduces new features that quickly become intractable, such as the mutual long-range repulsion among vortices and the many pinning configurations for the vortex array. In the transition to large numbers, much of the intuition developed for small numbers is lost. Carrying a lossless current, for example, requires all the vortices to be fully trapped by pinning defects. If significant numbers of vortices move, they dissipate energy and a finite resistance appears. The pinning force of one columnar pinning defect on one vortex line can be calculated and one could estimate the critical current a macroscopic array should support. Surprisingly, large arrays of vortice attain only 20–30% of the maximal theoretical critical current, which is given by the depairing current density, $J_{dp}$. The origin of this 'glass ceiling' for the critical current is not known, but presumably it reflects low energy mechanisms for vortex motion in large arrays that do not occur in small arrays. It tells us that small-array theory and simulations are not good guides for large-array behavior.

The issues that emerge for large-scale arrays include the following:

(i) The mutual vortex-vortex long-range repulsion allows the motion or trapping of a given vortex to influence its neighbors.

(ii) The flexibility of the vortex lines allows them to simultaneously seek and attach to many pinning defects at different locations and to move in intricate dynamic configurations as vortex segments detach from one pinning defect and attach to another.

(iii) The interruption of superconducting current pathways by non-superconducting pinning defects reduces the cross-sectional area available for supercurrent flow.

(iv) The possibility of vortices cutting and reconnecting in new configurations during their motion [7–9].

Although predicting the behavior of large-scale arrays of vortices is a fundamental challenge of high practical value, it has remained out of reach of analytical theory and conventional numerical simulation. With recent advances in high-performance computing, the time-dependent Ginzburg-Landau (TDGL) formulation of vortex behavior has come to the fore to capture the dynamics of large-scale arrays of vortices. The TDGL equations describe the spatial evolution of the complex superconducting order parameter and do not require tracking the motion of each individual vortex line. Instead, the TDGL formalism treats the superconductor as a continuous field describing the superconducting medium, punctuated by an array of singularities representing the vortices that arise spontaneously in the presence of a magnetic field. Within this framework, arbitrary arrays of pinning defects can be distributed throughout the superconducting medium, which attract vortices by lowering their free energy if their cores overlap with the pinning defects. The attractive pinning potential is widely tunable by adjusting the size and shape of the pinning defects. The TDGL equations automatically take into account the flexibility of vortex lines, long-range mutual vortex repulsion, vortex cutting and reconnecting, and the interruption of current paths by replacement of current carrying superconductor with insulating pinning defects. The challenge of this appealing TDGL formulation of vortex dynamics is the powerful computational platform needed to represent a macroscopic array of vortices and pinning sites. Earlier TDGL formulations of vortex dynamics were restricted to modest sized systems, where the full macroscopic behavior did not emerge. The recent development of scalable algorithms for leadership-class computing platforms have brought the numerical exploration of superconductors to a new level where direct experimental validation of finite sized samples can be realized with advanced large-scale, parallel solvers of the TDGL equations [6]. Fully macroscopic systems of vortices and pinning defects, can now be treated, allowing systematic investigations of the critical current, in particular how it is affected by temperature, magnetic field strength and orientation, and the size, shape and concentration of pinning defects. TDGL simulations have the advantage of high versatility and are applicable not only to copper oxide high-temperature superconductors — now the most promising for applications — but also to any other superconducting family by simply adjusting the materials dependent phenomenological parameters of superconducting coherence length, penetration depth, vortex elasticity and transition temperature. A brief summary of the TDGL formalism is presented later.

Below, we highlight some of the recent progress in enhancing the current carrying capacity of commercial high-temperature superconductors and the insights that TDGL formulation has elucidated. These include (i) the potential maximum critical current that can be carried by a superconductor with a large array of vortices and pinning defects, (ii) the optimal size, shape, and concentration of pinning defects that achieve the maximum current, (iii) how these parameters vary with temperature, the magnitude and the direction of the magnetic field, and the size of the system, and (iv) what novel genetic algorithms can tell us about defect evolution and vortex pinning. Answering these questions in the context of a versatile TDGL simulation validated with controlled experiments represents a major step forward for the fundamental understanding of the dynamics of macroscopic arrays of vortex lines and for the practical development of high-critical-current superconductors for targeted applications.

# Critical current in high-temperature superconductors

In contrast to conventional Nb-based superconductors, high-temperature superconducting cuprates are not amenable to the traditional approaches of alloying and wire drawing. Furthermore, the detrimental effects of grain boundaries in cuprates [10–12] necessitate new conductor architectures. A conductor layout that is scalable to industrial levels and that has yielded the highest $J_c$ to date is the so-called coated conductor (CC) geometry schematically shown in Fig. 2. Various industrial approaches, which differ in substrate material and substrate treatment, composition of buffer layers and superconductor deposition methods, have been demonstrated [13]; however, all share the overall rationale to grow essentially epitaxial superconducting films on flexible metallic substrates. Most work on coated conductors has been devoted to YBCO, or more generally REBCO (Y → RE), conductors [13–16].

'Pristine' REBCO films contain a variety of naturally occurring defects that can serve as vortex pin sites [57] such as point defects, dislocations, stacking faults or twin boundaries. However, the critical currents produced by these naturally occurring defects are not sufficient for most applications. Additional pin sites are typically introduced into the REBCO matrix by tuning precursor chemistry and growth conditions such that non-superconducting precipitates of various sizes and shapes, so-called artificial pinning centers, appear during the growth process that provide strong pinning by so-called artificial pinning centers [18–20]. Significant enhancements of $J_c$ have been achieved with the incorporation of $RE_2O_3$ [21, 22], $YBa_2Cu_3O_5$ [23], $BaZrO_3$ [24–26] nanoparticles. A major advance in the synthesis of high-performance REBCO coated conductors was the discovery that under certain deposition conditions the addition of excess metal oxides such as $BaZrO_3$ [27–37], $BaSnO_3$ [38] or $BaHfO_3$ [39] results in the formation of self-assembled nanorods largely oriented along the c-axis of the YBCO structure with lengths of hundreds of nm and diameters of ~ 10 nm. Their geometry makes them ideal pinning sites, particularly in aligned magnetic fields, and indeed, in recent years, remarkable advances in performance of research-scale REBCO CCs has been achieved by increasing the Zr-content and perfecting the structure of the BZO nanorods [32–37]. In short-length samples of highly doped REBCO conductors, critical current densities as high as 7 MA/cm$^2$ at 30 K and 9 T applied parallel to c-axis have been reported [34–36].

Ultimately, the figure of merit for applications is the whole-wire or engineering critical current density, $J_e$. Since the thickness of the REBCO layer in typical commercial CC is around 1 µm while the entire conductor thickness is in the range of 50–100 µm, $J_e$-values are significantly smaller as compared to $J_c$. Attempts to grow thick CC have encountered the problem of reduced average $J_c$ arising due to the deterioration of the microstructure with increasing thickness [18, 40, 41]. Only recently, it has been possible to achieve engineering critical current densities above 5 kA/mm$^2$ at 4.2 K and 14 T || c in short-length CC of > 4 µm thickness [42, 43]. Such remarkable in-field performance of REBCO conductors opens a route for increasing the current carrying capacity by increasing the thickness of the coated conductor and bodes well for high-field magnet applications.

The synthesis of desired pinning structures such as nanorods requires carefully tuned chemistry and growth parameters. However, their stable implementation for industrial fabrication of long-length thick coated conductors has proven a challenge. Furthermore, the growth of nanorods is not a practical option for solution-grown REBCO films. An alternative to modifying the chemical synthesis in order to generate the desired pinning structure is afforded by particle irradiation, which is a generic technique applicable to all superconducting materials that are sufficiently thin (typically tens of micrometers). Depending on the mass and energy of the ions and the properties of the target material, irradiation enables the creation of defects with well-controlled density and topology, such as points, clusters or tracks [2]. Irradiation of superconductors to induce tailored defects to raise their critical current has so far remained in the realm of fundamental research. Only recently has the potential for improving the performance of coated conductors through particle irradiation been recognized [44–56].

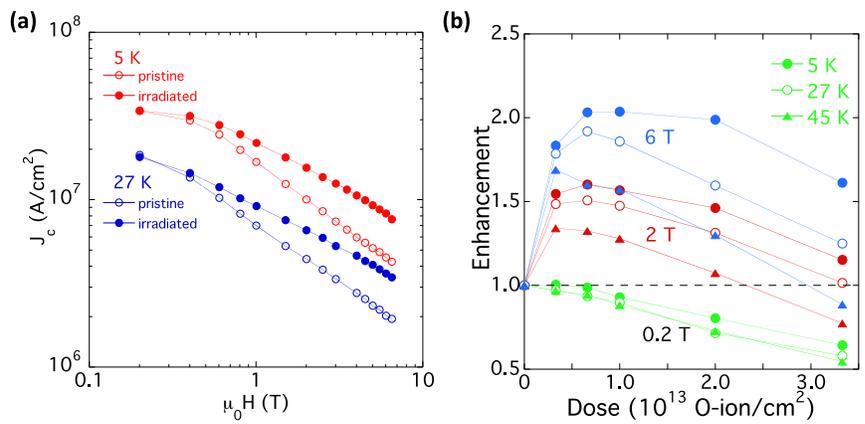

Figure 3. (a) Magnetic field dependence at 5 K and 27 K of the critical current density of superconducting coated conductor samples from AMSC both before and after oxygen irradiation at 3.5 Mev. (b) Irradiation dose dependence of the enhancement of the critical current at temperatures of 5, 27, and 45 K (identified by open and closed circles and closed triangles, respectively) in fields of 0.2, 2 and 6 T (represented by green, red, and blue colors, respectively).

While state-of-the-art commercial HTS wires have been engineered to optimize their current carrying capacity over the course of more than a decade of research, recent studies have demonstrated a new method to enhance the critical current of these coated conductor wires at high magnetic fields with ion irradiation [45]. Initially, 4 MeV proton irradiation to a fluence of $8 \times 10^{16}$ p/cm$^2$ were shown to double the critical currents in fields of 6 T at $T$ = 27 K on short samples cut from production-line HTS coated conductors. This is a field and temperautre range of interest for application in rotating machinery. Further studies demonstrated that 3.5 MeV oxygen irradiation through the protective silver coating (see Fig. 2) of a commercial second generation superconducting tape could also reproduce a doubling of the critical current but at an exposure of just one second per 0.8 cm$^2$ [52]. Transmission electron microscopy studies on these samples showed a mixed pinning landscape [Fig. 3(b)] comprisied of pre-existing precipitates and twin boundaries and small, finely diespersed irradiation induced defects of ~ 5 nm typical dimension, which are particularly effective for vortex pinning in high fields. Significant effort has been devoted to reducing the size of artificial pinning centers to the 5-nm level by chemical synthesis, but this requires modifications of the growth process and remains a challenge. In contrast, irradiation can easily produce defects of this size and is independent of the growth process. Furthermore, particle irradiation can create defect structures that are highly uniform along the conductor length, [51] an aspect that has proven to be a challenge for chemical synthesis routes. Later, we demonstrate how controlled irradiation and TDGL can be used synergistically to elucidate vortex dynamics and to validate and predict the effectiveness of certain types of defects.

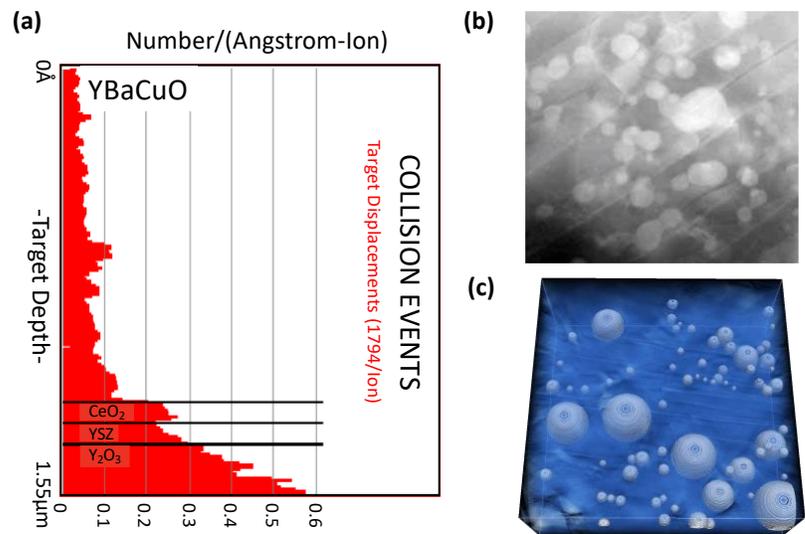

Figure 4. (a) TRIM-SRIM simulations of collision event as a function of depth. (b) 2D scanning transmission electron microscopy image at different angles. (c) 3D reconstructed landscape.

Here, SRIM-TRIM simulations [57] were used to tune the irradiation in such a way as to achieve high and uniform damage in the REBCO layer [Fig. 4(a)]. The enhancement of the critical current as a function of field and dose is shown in Figs. 3(a) and 3(b). Such studies have led to the realization of potential high-speed industrial reel-to-reel post-processing methods using heavier ions [51].

# Time-dependent Ginzburg-Landau model for pinning in type-II superconductors

The vortex dynamics in strong type-II superconductors can be described by the TDGL equation, which has the form $(\partial_t + i\mu)\psi = \varepsilon(\mathbf{r})\psi - |\psi|^2\psi + (\nabla - i\mathbf{A})^2\psi + \zeta(\mathbf{r}, t)$ in dimensionless units. Here $\psi(\mathbf{r}, t)$ is the complex order parameter describing the superconducting wave function, $\mu = \mu(\mathbf{r})$ is the electric scalar potential, $\mathbf{A}$ the vector potential associated with the external magnetic field $\mathbf{B} = \nabla \times \mathbf{A}$, and $\zeta(\mathbf{r}, t)$ is a temperature-dependent noise term. The unit of length is the superconducting coherence length $\xi = \xi(T)$ at a given temperature $T$, and the unit of magnetic field is the upper critical field, $H_{c2} = H_{c2}(T) = \hbar c/2e\xi^2$. Using this description, one obtains the supercurrent in the system as $\mathbf{J}_s = J_{dp} \text{Im}[\psi(\nabla - i\mathbf{A})\psi]$ and the normal current $\mathbf{J}_n = -J_{dp}\nabla\mu$, where $J_{dp}$ is the depairing current. With this definition of the current, we can then obtain the critical current by solving the TDGL equations numerically [6, 58] and using a voltage threshold criterion. Defects in the system are modelled as so-called $\delta T_c$-defects [2], meaning that the critical temperature, $T_c$, is spatially modulated such that $\varepsilon(\mathbf{r}) \propto [T_c(\mathbf{r}) - T]$. In many cases, defects are modeled as non-superconducting ellipsoidal inclusions with a reduced critical temperature

# Strong pinning by extended defects

The pinning behavior of randomly placed spherical defects which can be realized in real samples by self-assembled nanoparticle inclusions or oxygen irradiation was studied using the numerical approach described above and analytical considerations [59, 60]. Here, one has to distinguish between small and large defects, as the resulting critical currents show fundamentally different behavior in larger fields. In the case of a small density of smaller defect particles, the vortex lattice preserves its structure and the critical current density $J_c$ decays with the magnetic field following a power-law $B^{-\alpha}$ with $\alpha \cong 0.66$, which is consistent with predictions of strong-pinning theory, see Fig. 5(a). For a higher density of particles and/or larger inclusions, the lattice becomes progressively more disordered (Fig. 6) and the exponent smoothly decreases down to $\alpha \cong 0.3$, see Figs. 5(a) and 5(b). At high magnetic fields, all inclusions capture at least one vortex and the critical current decays faster than $B^{-1}$ as one would expect from theoretical considerations. In the case of larger inclusions with a diameter of four coherence lengths, $4\xi$, the magnetic-field dependence of the critical current is strongly affected by the ability of inclusions to capture multiple vortex lines. At small densities, the fraction of inclusions trapping two vortex lines rapidly grows within a narrow field-range leading to a peak in $J_c(B)$-dependence within this range, see Figs. 5(b). With increasing inclusion density, this peak transforms into a plateau, which then flattens out. The peak position (as a function of the magnetic field) depends only on the defect size and is therefore distinct from the classical peak effect arising near $H_{c2}$ due to a softening of the vortex lattice elasticity.

In the intermediate magnetic field regime where the vortex lattice is disordered and a finite fraction of inclusions is occupied with vortex lines, the critical current decays with the magnetic field as a power-law $B^{-\alpha}$, where the exponent $\alpha$ decreases with increasing inclusion density (for diameter $2\xi$ it drops from 0.66 to 0.3). Here the exponent $\alpha$ is mostly determined by the volume fraction occupied by the inclusions. At a certain magnetic field, which depends on the inclusion size, all inclusions become occupied by at least one vortex and the critical current is roughly inversely proportional to the field. However, above this field, the critical current decreases faster than the $B^{-1}$ dependence, since vortex-vortex interaction and multiple occupation of defects effectively reduce the pinning forces.

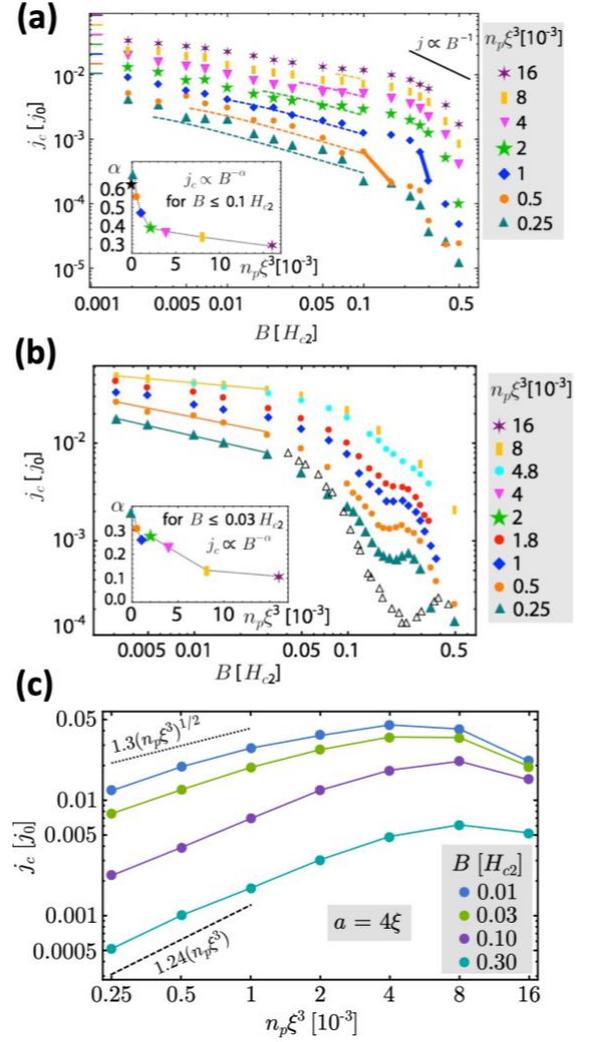

Figure 5. Critical current behavior in the strong pinning regime: (a) Decrease of the critical current with magnetic field for small spherical pinning sites (diameter $2\xi$) for different defect densities ($n_p$) occupied by all defects. Inset shows the scaling exponent as function of defect density. (b) The same as (a), but for larger defects (diameter $4\xi$): At high fields, defects can be multiply-occupied, resulting in a local peak in $J_c(B)$. Open triangles show single defect results. (c) Non-monotonic dependence of the critical current as function of volume fraction in different fields for $4\xi$-defects. Results agree with oxygen irradiation studies.

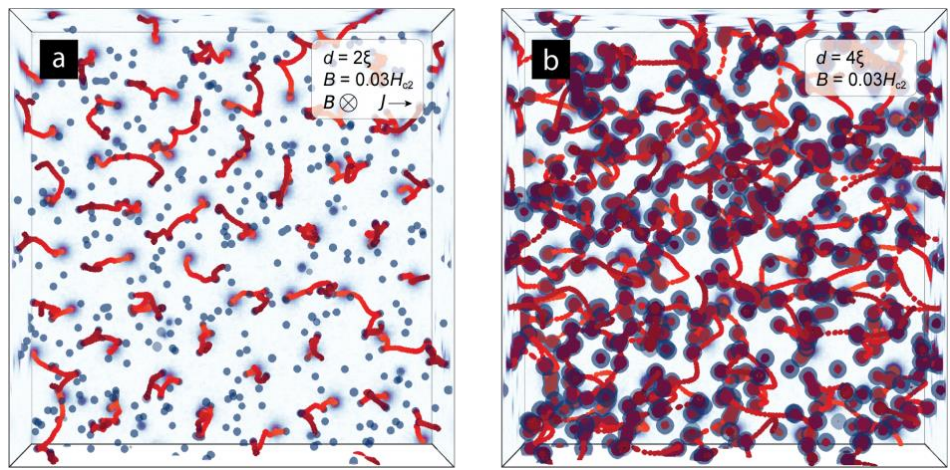

Figure 6. Vortex configurations in the strong pinning regime: (a) For spherical defects of diameter $2\xi$ (b) for defects of diameter $4\xi$.

For diameters $2\xi$ and low inclusion densities $n_p$, the lattice becomes ordered at a magnetic field which rapidly increases with $n_p$. The ordering transition — driven by increasing the magnetic field — is accompanied by a reduction of the particle fraction occupied by vortex lines and by a sharp drop of the critical current.

## Validation on real pinning landscapes

Clearly, it is important for numerical calculations of critical currents to be validated with realistic scenarios. Such validattion was obtained in a detailed study of the vortex dynamics in a 'real' pinning landscape that was simulated using TDGL. The defect distribution was captured by imaging information of pinning centers obtained by scanning transmission electron microscopy tomography of Dy-doped $YBa_2Cu_3O_{7-\delta}$, which allowed for the reconstructing of the real three-dimensional pinning landscape of the experimentally measured sample. This information was subsequently used to model the exact same landscape as in the experimental system — down to the resolution limit of the tomography, using almost the same size as the sample. It was shown that the simulated magnetic field dependence of the critical current is in good qualitative and nearly quantitative functional agreement with the experimentally observed behavior. This work also suggests that more detailed scanning transmission electron microscopy tomography studies could further improve the quantitative agreement — in particular if smaller defects could be detected, which are introduced by higher Dy doping. Overall, Ref. [61] clearly shows that TDGL simulations are a valuable tool, faithfully describing the dynamic properties of high-performance type-II superconductors with the potential for predictive capabilities and analysis.

Further evidence for the validity of the TDGL approach was obtained in Refs. [3, 62], where experimental systems — in these cases irradiated REBCO tapes and patterned superconducting films — were compared to simulated systems.

# Critical current by design

Following the above experimental validation studies, systematic TDGL simulations were used to determine the optimum vortex pinning landscape for randomly distributed metallic spherical inclusions [63]. The simulations allow the prediction of the size and density of particles for which the highest critical current is realized. Using a simulated system size of $100\xi \times 100\xi \times 50\xi$ where $\xi$ is the superconducting coherence length, with $256 \times 256 \times 128$ mesh points and periodic boundary conditions in all directions, the simulations revealed that for a given particle size and magnetic field, the critical current reaches a maximum value at a particle density that corresponds to about 15–23% of the total volume being replaced by the nonsuperconducting material, see also Fig. 5(c). Furthermore, for a fixed diameter, this optimal particle density increases with the magnetic field. Moreover, as the magnetic is field increased, the optimal particle diameter slowly decreases from 4.5 to 2.5 coherence lengths.

Large-scale TDGL simulations that automatically include all mesoscale vortex interactions while tracking the vortex dynamics have also been used to demonstrate the competing effects of vortex pinning in a mixed defect landscape. Fig. 7(a) depicts the angular dependence of the critical current at a fixed field and temperature of a commercial HTS coated conductor tape from SuperPower Inc. These conductors contain self-assembled nanorods of BaZrO that act as strong vortex pinning centers aligned normal to the substrate of the conductor. Hence, a maximum in $J_c$ is reached when the applied magentic field is oriented $B || c$ (normal to the tape), whereby the the vortices are aligned with and pinned by the nano-rod. A subsequent irradiation with high-energy Pb-ions at 45° from the normal creates a set of columnar defects. Similar to the nanorods, one may expect to see two peaks in the $J_c$, one at 0° due to the nanorods and another at 45° from vortex pinning by the columnar defects. Surprisingly, only one peak due to the columnar defects is observed in the blue curve of Fig. 7(a). TDGL simulations were able to reconstruct the qualitative behavior of $J_c(\theta)$ and revealed an additional sliding motion of vortices along the columnar defects which was responsible for the reduction of the $J_c$ peak at 0° (see Fig. 7 lower panels) [3]. These studies directly demonstrate the non-additive nature of vortex pinning in a mixed defect landscape. Indeed, Fig. 8 shows that the addition of columnar tracks to the YBCO sample with pre-exisiting BZO nanorods may both increase and decrease the magnitude of the critical current density. For external field $B$ at $\theta = 45°$ to $c$-axis $J_c$ increases at low matching field $B_{\Phi,C}$ (i.e. low density) of the tracks, reached the maximum at some value of $B_{\Phi,C}$, and then decreases with further increase of $B_{\Phi,C}$. Yet another evidence of strong non-additivity of different defects is the suppression of the pinning potential of the superconductor-vacuum interface due to additional defects near this interface [64]. In the same work it was also shown that the optimal landscape can be achieved for a certain gradient in the defect density. In general, the consideration of geometrical pinning due to boundaries in mesoscopic superconductors, where both bulk and surface effects are important, makes the prediction of critical currents even more challenging.

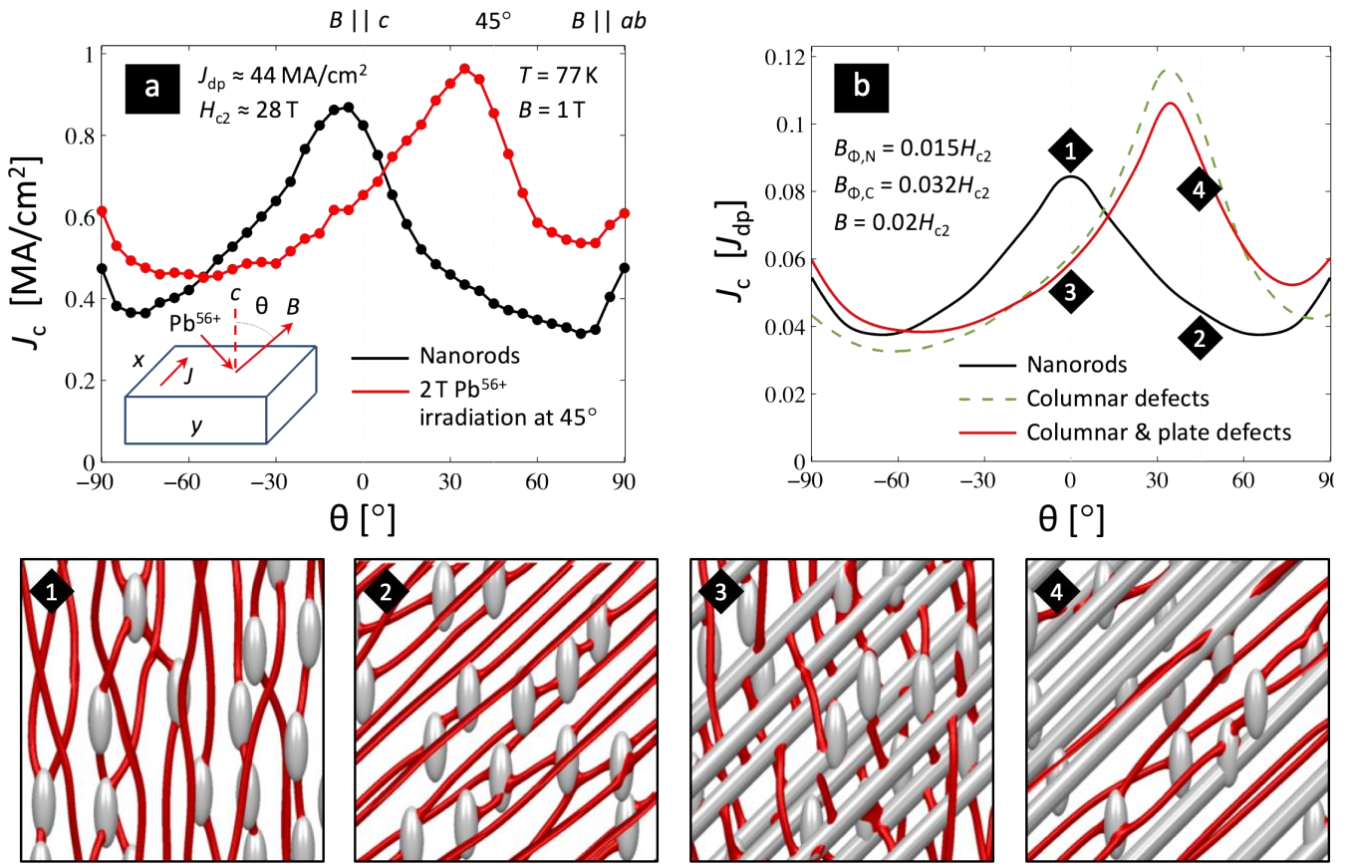

Figure 7. Non-additive nature of vortex pinning. (a) Angular dependence of the critical current for a pristine YBCO coated conductor containing BaZrO self-assembled nanorods and the same conductor after irradiation with 1.4 GeV Pb-ions at a 45 degree angle, (b) TDGL simulations at various fields for the two cases in panel (a). Isosurfaces of order parameter in TDGL simulations corresponding labeled by diamonds 1-4.

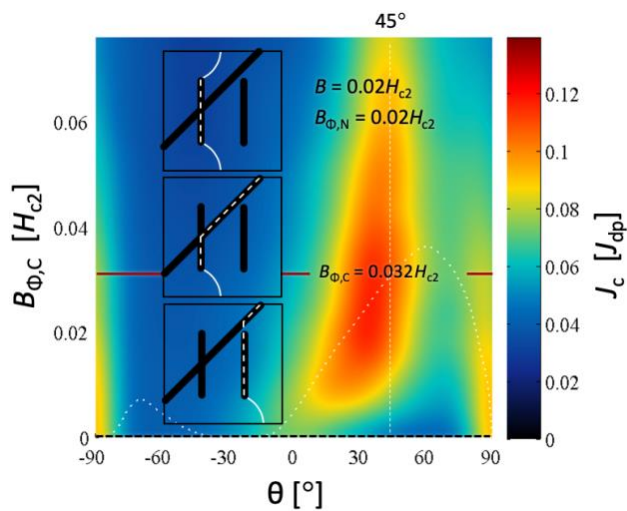

Figure 8. Angular dependence of the critical current for different concentration of columnar defects $B_{\Phi,C}$ tilted at $\theta = 45°$ and at a fixed concentration of nanorods. With increasing concentration of columnar defects, the peak shifts from $\theta = 0°$ to 45°. The near-optimal critical current corresponds to the angular range $\theta = 30–40°$ and concentration of columnar defects $B_{\Phi,C} \approx 0.02–0.04 H_{c2}$ (or volume fraction = 6–11% occupied by them). Inset: Vortex sliding on tilted columnar defects.

## Optimization of the defect size and density

A similar idea also works for other types of defects. For instance, by means of randomly placed monodisperse spherical defects one can model some chemical self-assembled defects, as well as defects arising during irradiation with light ions such as oxygen. The small size and low concentration of the defects provide insufficient pinning for the vortex lattice and thus generate low $J_c$. High concentrations of large defects reduce the effective potential for pinned vortices, so they can easily jump from defect to defect, which in turn also produces a small $J_c$. As indicated before, the maximal $J_c$ for spherical defects is achieved when those have a diameter $d = 4\xi$ for external magnetic field $B = 0.05 H_{c2}$ and $d = 2.5\xi$ for $B = 0.2 H_{c2}$ [59], where the external field is applied at a fixed angle, perpendicular to the applied current. It turns out that the volume fraction occupied by defects, $f \approx \rho \pi D^3/6$, is typically a more practical quantity for the optimization than the defect density, $\rho$, itself. Utilizing this definition, it was found that the optimal volume fraction is $f \approx 20\%$ and nearly field independent. However, in general, there is no universal optimal pinning configuration for all magnetic fields. Thus, for best performance in a given application, pinning landscapes should be designed by taking into account the relevant magnetic field range.

Analogous optimizations were performed for (i) randomly placed identical spheroids and (ii) combination of randomly placed for spherical and columnar defects in Ref. [65]. It follows that in both these cases, the optimal pinning landscape consists of columnar defects along the magnetic field. Table 1 shows the comparison between different optimal pinning landscapes in the same field.

| Pinning landscape | $J_c/J_{dp}$ | Pinning landscape parameters | Robustness of the optimal configuration | Ref. |
|---|---|---|---|---|
| Randomly placed spheres | 6.1% | Volume fraction, $f$ = 22%, defect diameter, $d$ = 3.5$\xi$ | Shallow peak both in $J_c(B)$ and angular dependence of $B$. Shallow peak in $J_c(f)$ and $J_c(d)$ | [59, 65] |
| Randomly placed spheroids \|\| $B$ and $\perp J$ | 9.0% | Volume fraction, $f$ = 20%, diameter $\perp B$, $d$ = 3.0$\xi$, diameter \|\| $B$, $d_{\|\|}$ = $\infty$ | Shallow peak in $J_c(B)$; robust against rotation, φ, around $B$; not robust against rotation, θ, around $J$. Shallow peak in $J_c(f)$ and $J_c(d)$ | [65] |
| Randomly placed spheres and columns \|\| $B$ and $\perp J$ | | Spheres volume fraction = 0, columns volume fraction, $f$ = 20%, columns diameter $\perp B$, $d$ = 3.0$\xi$ | | [65] |
| Columns \|\| $B$ and $\perp J$ arranged in hexagonal pattern | 32% | Hexagonal lattice constant, $l$ = 8.5$\xi$, defect diameter, $d$ = 3.5$\xi$ | Narrow peak in $J_c(B)$ and $J_c(\theta)$, shallow peak in $J_c(\phi)$. Narrow peak in $J_c(l)$, shallow peak in $J_c(d)$. Robust against small random displacements | [63] |
| Equidistant planar defects \|\| $B$ and \|\| $J$ | 40% | Distance between defects, $h$ = 7.4$\xi$, defect thickness $d \approx 0.5\xi$ | Shallow peak in $J_c(B)$, narrow peak in $J_c(\phi)$ and $J_c(\theta)$. Shallow peak in $J_c(h)$ and $J_c(d)$ | [63] |

Table 1. Optimal critical current $J_c$ [% of the depairing current $J_{dp}$] for the same external field of $B$ = 0.1$H_{c2}$ for different pinning landscapes.

## Evolutionary approach targeting critical currents

It can be seen from the above that changes in sizes and concentrations can affect the superconducting current to varying degrees. Up to this point, only sets consisting of identical defects have been considered. Real superconductors consist not only of different types of defects, but the sizes and shapes of defects may vary within a single type. Moreover, only certain types of defects may be added/changed experimentally. How can one effectively describe defects with a spread of parameters? If this spread is small, it is natural to introduce randomness in such parameters with some certain distribution. However, in real superconductors, this distribution can be highly nontrivial [66]. The most reliable way to describe and improve such complex systems is to consider each defect individually.

The ability to systematically characterize and improve superconductors with different defects is driven by access to significant computing power and the development of sophisticated computer algorithms. Modern computer systems have enabled the application of *genetic algorithms* in electromagnetic design of space antennas, financial mathematics and energy applications, among others. To implement genetic algorithms, an objective function which needs to be minimized or maximized is required. For example, it can be the maximization of the emitted/absorbed electromagnetic radiation of the antenna or the minimization of risks in financial models. Within this approach, one only needs to specify the direction of positive mutations in order to optimize the characteristics of the system of interest or more precisely, replace natural selection with targeted selection. Targeted evolution is especially effective in complex systems with many degrees of freedom.

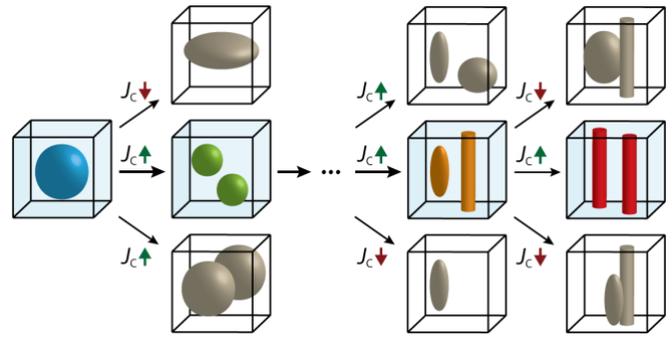

Figure 9. Sketch of the targeted evolution. Algorithm rejects negative mutations decreasing critical current and accepts positive mutations increasing critical current. Each mutation may vary size, shape, or position of each defect individually as well as add or remove defects. The evolution ends at the configuration with maximal critical current (red) in the last generation [63].

This kind of genetic approach can be applied to increase critical current in superconductors having various defects inside by setting the critical current density, $J_c$ as the objective function. The approach is to track the dynamic transformation of the shape, size and multiplicity of a particular defect as it evolves and interact with vortex matter to generate the highest critical current density, see schematic Fig. 9. The genetic algorithm carried out for low anisotropy and metallic defects in Ref. [63] demonstrated that in the case of a fixed direction and magnitude of the magnetic field $B$ and a fixed direction of the external current $J$ ($J \perp B$), the pinning configuration producing the highest critical current consists of planar defects parallel both to $J$ and $B$, see Table 1 and Fig. 11(a). A typical order parameter amplitude distribution in this configuration is shown in Fig. 10(b) in contrast to a hexagonally ordered array of columnar defects shown in Fig. 10(a).

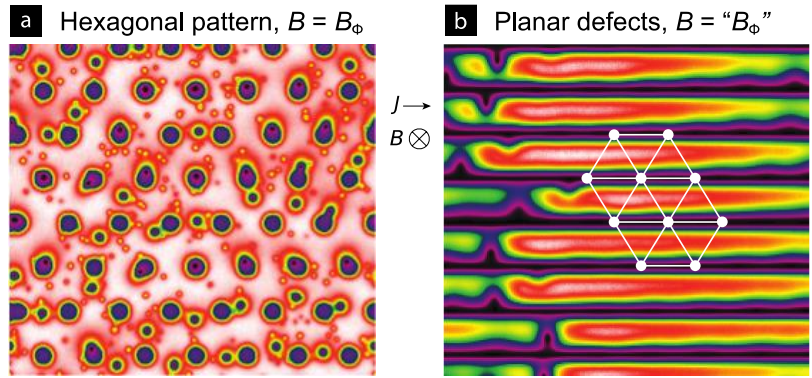

Figure 10. (a) Order parameter amplitude for hexagonal array of columnar defects. (b) and for planar defects at the matching field.

Generally, from the $J_c$ values summarized in Table 1, one can conclude that:

(i) Optimal pinning landscapes strongly depend on the range of the external magnetic field and on the mutual orientations of the applied current and magnetic field. Typically, higher $J_c$ can be achieved with more restrictions in orientations of the field and current.
(ii) Ordered lattices of defects are much better than random placement. Typically, small deviations in the optimal positions of the defects do not reduce $J_c$ significantly.
(iii) Highest critical currents (in high external fields) are reached only in the collective pinning/depinning regime. If vortices are able to jump from defect to defect one-by-one, it is a clear indication that one can further improve the pinning landscape.

An important feature of well optimized pinning landscapes is its extremely sharp *I-V* curve. For example, the *I-V* curve for optimized planar defects [63] is much sharper than the *I-V* curve for randomly placed spherical defects [59]. This suggests that the best pinning landscapes have no 'smeared', low resistive region in the I-V curve below the maximum possible value of $J_c$, which is probably related to localized vortex motion. As a result, when this critical current is exceeded, the system immediately turns into a highly resistive state due to collective, avalanche-like depinning, which in turn may lead to overheating.

The genetic algorithm can also be applied to pre-existing defects in superconductors, and used to predict the morphology of 'additional' defects that can be compounded to the already existing defects in a commercial superconductor. Such a procedure can help with the improvement of existing commercial HTSs, see Fig. 11. Here, one initializes the pinning landscape with artificial fixed pre-existing defects, which blocks to some extent the supercurrent flow. In Fig. 11(b) for example, a set of equidistant tilted planar defects was added to the system periodic in all directions, forcing the current to flow at an angle and the (free) solution shown in Fig. 11(a) does not work anymore. Similarly, in Fig. 11(c) two semi-ellipses were included to the system with open boundary conditions in the direction of the Lorentz force (vertical direction in the figure). These pre-existing defects cannot be modified by evolution and significantly change the optimal pinning landscape corresponding to nearly-optimal critical current. In both cases the total optimal critical current is somewhat smaller than in the case of absent pre-existing defects, but nevertheless, enhances the critical current over the one with only pre-existing defects only by ~ 2.6 times in scenario b [shown in Fig. 11(b)] and 3.0 times in scenario c [Fig. 11(c)] [63].

In Fig. 11(d) the current can be applied in both horizontal and vertical directions. The objective function is min{$J_{c\rightarrow}$, $J_{c\uparrow}$}, where $J_{c\rightarrow}$ is horizontal critical current and $J_{c\uparrow}$ is vertical critical current. The evolutionary algorithm with this objective function results in a set of hyperuniformly ordered [62, 67] columnar defects along the external magnetic field. The corresponding critical current is $J_c = 0.267 J_{dp}$ is close to the critical current density for an optimized hexagonal lattice [63].

Overall, the combination of the possibility to quantitively calculate critical currents numerically and genetic algorithms allows the systematic prediction of modifications that can lead to better critical current density and can potentially replace the traditional trial-and-error approach to defect design. Furthermore, this targeted evolution of defect morphology via simulations can be coupled to guided synthesis approaches (described below), which together herald a new approach for faster optimization of highly complex defect landscapes and 'designing critical currents'.

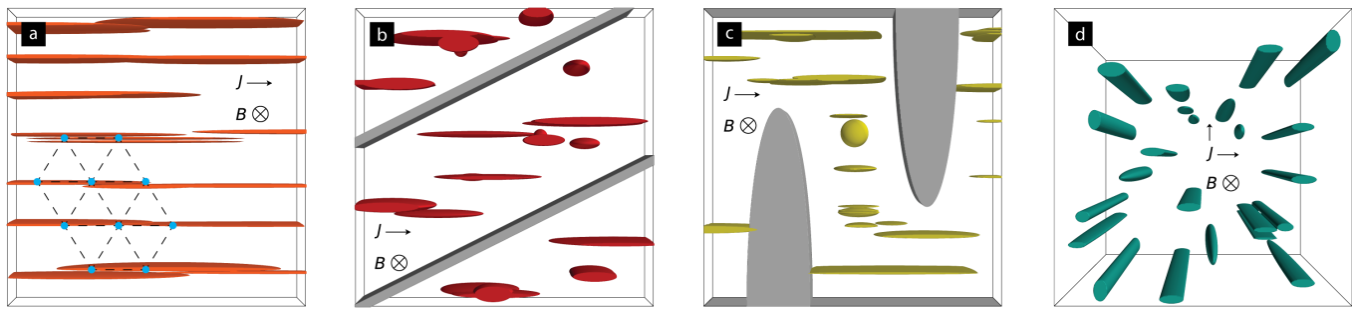

Figure 11. Different scenarios for evolution algorithm [63]. (a) Algorithm having flexibility to modify all defects in the system resulting in the pinning landscape consisting of almost equidistant planar defects. (b, c) Evolution of the pinning landscapes with for predefined environments (tilted planar defects in b and two half-ellipses shown in grey). The nearly optimal pinning centers are shorter planar defects parallel at the same time to the applied current and external magnetic field. (d) In the last scenario, the current direction is not fixed anymore, but is applied in two directions perpendicular to magnetic field. The pinning landscape evolves to hyperuniformly placed columnar defects.

## Computation-assisted material synthesis

In addition to guiding the identification of optimal pinning configurations as described above, large scale similations are also poised to advance the *materials synthesis* of such structures, in particular of self-assembled defect structures. Since the lattice mismatch between self-assembled BZO nano-rods and the REBCO matrix is large, ~ 8%, lattice strains are an important component in the energetics of the system. In fact, extensive work on solution-grown YBCO films has shown that strain-fields associated with self-assembled nano-particles can give rise to strong isotropic pinning [25, 68]. Continuum elasticity theory [69] has successfully been applied to establish macroscopic parameters influencing the diameter and separation of self-assembled nanorods while other features such as the length and splay of the nano rods, or the role of stacking faults in reducing strains are currently not well understood. One can expect that a realistic model description of the nanorod/matrix morphology based on a combination of discrete atomistic and continuum elasticity approaches with experimental input parameters will soon be enabled by large-scale simulations. These simulations could elucidate the role of mismatch strain between the REBCO matrix and the nanorods, the incoherent strain in the REBCO matrix and the effect of stacking faults on strain release. In fact, a recent three-dimensional Monte-Carlo (MC) simulation [70] of a simple model system containig two types of cubic materials unit cells and taking into account only adsorption/desoorption and diffusion along the surface has yielded the formation of self-assembled nanorods for certain values of deposition rates and temperatures. These type of simulations can provide a first insight into the factors determining the length, splay and concentration of defects and are therefore a good starting point to understand the actual chemical vapor deposition process. For more realistic simulations, the diffusive dynamics of the building blocks at the surface should be taken into account by, e.g., molecular dynamics (MD) simulations. Therefore, the inclusion of MD simulations can help to predict lattice strains caused by self-assembled nanoscale inclusions. Careful validation and estimation of parameters and interaction potentials is essential for realistic modeling and requires large sets of experimental results for comparison. Hence, as with large scale TDGL simulations, the MC and and MD approach may provide a rational route to optimize the synthesis of effective defects in these high temperature superconductors.

## Summary

Currently, in its present form, critical-current-by-design exploits large-scale, time-dependent Ginzburg-Landau simulations that automatically include all mesoscale vortex interactions and quantitatively predict the resultant macroscopic critical current. However, its potential may be even more far-reaching. For example, TDGL coupled with heat diffusion equations will allow treatment of thermal breakdown effects in high-current or field applications, providing *predictions* of magnet quench and catastrophic thermal runaway conditions and ways to mitigate them. Furthermore, incorporating the Maxwell equations within the TDGL equations to handle magnetic dipoles and accurate boundary conditions with external fields can guide novel magnetic pinning strategies for mitigating dissipation in the ubiquitous vortex liquid state in HTS. As multi-band iron-based superconductors have emerged as a potential for cost-effective next generation HTS wires, the TDGL equations can be adapted to multi-band situations by introduction of multi-component order parameters. Moreover, integrating the critical current simulations with the combined MC/MD approach for defect synthesis provides a complete toolkit for the in-silico design of pinning landscapes in high-performance superconductors. Finally, using the emerging state-of-the-art machine learning and artificial intelligence algorithms will allow in the future to automatically predict not only the best pinning landscapes for a given application, but also the protocol on how to synthesize these landscapes and to determine the resulting critical current. Here, machine learning algorithms can be used to improve the fidelity of the TDGL and synthesis simulations by training the underlying model with extensive, existing experimental data set, consisting of transmission electron microscopy and x-ray diffraction information, and artificial intelligence to further adapt the targeted evolution in the above-mentioned genetic algorithms to predict arbitrary pinning landscapes more rapidly by, e.g., avoiding non-beneficial mutations or guiding the selection away from local optima.

The continuing advancement of powerful computers will enable further development of large-scale TDGL and MC/MD simulations, driving the critical-current-by-design paradigm to tackle more complex vortex pinning defect landscapes. It is already affecting the way we think about vortex pinning landscapes, replacing trial and error approaches with powerful simulations that have identified desirable defect morphologies for critical current enhancement for targeted applications. The critical-current-by-design paradigm can provide the basis of new design rules for optimizing critical currents in all applied superconductors.

# Acknowledgements

This work was supported by the U.S. Department of Energy, Office of Science, Basic Energy Sciences, Materials Sciences and Engineering Division.

The simulation results presented here are based on codes, which were developed within the Scientific Discovery through Advanced Computing (SciDAC) program OSCon funded by the U.S. Department of Energy, Office of Science, Advanced Scientific Computing Research and Basic Energy Science, Division of Materials Science and Engineering. We also thank Oak Ridge LCF, supported by DOE under contract DE-AC05-00OR22725, Argonne LCF (DOE contract DE-AC02-06CH11357), and the computing facility at Northern Illinois University, were many of the simulations were carried out.

Other simulations using time-dependent Ginzburg-Landau model can be found at [OSCon website](#) and [YouTube channel](#).

# Table of contents